# A stochastic approximation method for price-based assignment of autonomous EVs to Charging Stations

Georgios Tsaousoglou, Konstantinos Steriotis, Emmanouel Varvarigos
National Technical University of Athens

*Abstract*—**This paper proposes a method for the stochastic estimation of Charging Stations' prices based on dual decomposition. Prices are determined so as to minimize the social cost of the EVs while satisfying their constraints in expectation.**

## I. INTRODUCTION & RELATED WORK

In the near future, car-parking stations are expected to be replaced by Electric Vehicles (EV) parking stations that will also have the ability to charge the batteries of the EVs parked in them. We consider a certain geographical area containing a given set of Charging Stations (CSs). Users (EV drivers) wish to drive to a certain place (e.g., their work), park as near as possible to it, and, upon departure, receive their EVs charged to a predefined State Of Charge (SOC). Users generally prefer CSs close to their target location. If we let the system uncontrolled, the CSs located in "hot" areas will become congested, and users arriving at later times will be forced to use more distant CSs or find their batteries uncharged upon departure.

In this paper, we consider price-elastic users (users willing to park to a more remote CS in exchange for a lower price). Most previous works consider the self-optimization of a single CS [1], which is only part of the assignment problem considered here. In works where multiple CSs are modeled, the user's elasticity is considered either as known [2] or as a parameter that the user can report to the system if given the right incentives [3]. In contrast, we consider the problem where the user's elasticity is an intrinsic quality that is unknown and unobservable, even by the user himself. We tackle this problem by carefully assigning different prices to different EV-CS pairs via dual decomposition. We run Monte Carlo simulation scenarios to learn the appropriate multipliers (prices), given the probability distributions of the setting's parameters. In the proposed scheme, each user autonomously (through deliberate choice) picks his/her station upon arrival, while the overall allocation achieves a competitive ratio that is significantly better than the benchmark case. The proposed framework could be implemented in a mobile app, integrated with the user's GPS, so that users select their target location and are then presented with a set of nearby CSs along with their respective locations and prices. Thus, it comes out very natural to the end-user to pick a station intuitively.

## II. SYSTEM MODEL & PROBLEM FORMULATION

Let $N$ be the set of users requesting a spot in a certain area at a certain day, and $M$ be the set of CSs in that area. Continuous time is divided into $\tau$ time-intervals of equal duration. Set $T = \{1, 2, \dots, \tau\}$ contains the time intervals of the scheduling horizon.

Upon its arrival (EV entering the area) at time $a_i$, EV agent $i$ sends a message to the system containing: a) its target destination location $\mu_i$, b) its departure time $d_i$, c) the desired SOC $E_i$ at time $d_i$ and d) its charging rate capacity $e_i$. Let $T_i^f = \{t \mid a_i \leq t \leq d_i\}$ denote the set of timeslots during which EV $i$ is available for charging. Also, its target destination $\mu_i$ is modeled by its Cartesian coordinates $\mu_i = \{x_i, y_i\}$ on a 2D surface area. Finally, we conceptualize a parameter $\theta_i$, intrinsic to the user, which characterizes the disutility that the user experiences from parking away from his/her target location. More formally, each EV $i$ is characterized by its *type* $\psi_i$, which is the tuple $\psi_i = \{T_i^f, E_i, e_i, \mu_i, \theta_i\}$. The user will choose a CS to park/charge his/her EV. Let $\xi_{ij}$ denote the binary decision variable, indicating whether user $i$ chooses CS $j$. Since the user can only choose one CS, we have

$$\xi_{ij} \in \{0,1\}, \tag{1}$$

$$\sum_{j \in M} \xi_{ij} = 1, \quad \forall i \in N. \tag{2}$$

The CS $j$ is given the set $S_j = \{i \mid \xi_{ij} = 1\}$ of EVs that have chosen it, along with each EV's $T_i^f$, $E_i$ and $e_i$. Also, each CS $j \in M$ has a certain total energy $C_j^t$ available at timeslot $t$, representing energy it bought at the day-ahead market or its expected own RES production, similarly to [4]. $C_j^t$ is assumed known in this section, but it is later relaxed to follow a random distribution over a range of possible values. The CS agent of $j$ decides upon the state of each assigned EV $i$ at each timeslot $t$. This decision is expressed through indicator variable $u_{ij}^t \in \{0,1\}$, which is set it to 1 for charging, or 0 for not charging. The aggregate charging energy of EVs charging at CS $j$ in timeslot $t$ cannot be more than $C_j^t$ and an EV $i$ cannot be charged before arrival or after departure. Finally, the EV must acquire the desired energy until departure (assuming $E_i \leq (d_i - a_i) \cdot e_i$), exclusively from the CS that it chose:

$$u_{ij}^t \in \{0,1\}, \tag{3}$$

$$\sum_{i \in N} u_{ij}^t \cdot e_i \leq C_j^t, \quad \forall j \in M, t \in T^d, \tag{4}$$

$$u_{ij}^t = 0, t \notin T_i^f, \tag{5}$$

$$\sum_{t \in T} u_{ij}^t = \left\lceil \frac{E_i}{e_i} \right\rceil \cdot \xi_{ij}, \quad \forall i,j, \tag{6}$$

where $\lceil \cdot \rceil$ denotes rounding to the nearest higher integer. The location of each CS is fixed and thus the distance $l_{ij}$ of each CS from the user's target destination can be easily calculated. Let $L_i = \{l_{ij}, j \in M\}$ denote the set containing the distances $l_{ij}$ from the EV's target destination to each of the CSs. Ideally, the user would like to park as close as possible to his/her target destination. We model the user's disutility from parking away from his/her destination as the product of the squared distance $l_{ij}$ and the elasticity parameter $\theta_i$, which characterizes the user. Thus, our objective is to minimize the sum of users' disutilities

$$\min_{\xi_{ij}, u_{ij}} \left\{ \sum_{i \in N^d} \sum_{j \in M^d} \xi_{ij} \cdot \theta_i \cdot (l_{ij})^2 \right\}. \quad (7)$$
$$\text{subject to } (1) - (6).$$

The difficulty in problem (7) is threefold. First, it is an integer program for which an optimal solution may or may not be reached in a reasonable computational time. Second, the parameters $\theta_i$ are unknown and intrinsic to the users, meaning that $\theta_i$ is not only private to the user, but the users themselves cannot clearly define it and could not report their $\theta_i$ even if they wanted to. Third, the information about EV types is not known in advance but it is revealed in an online fashion.

Let us temporarily assume that the EV agents actually schedule their arrivals and needs from the day before. Even for the offline case, we still cannot solve Problem (7) because of the unknown intrinsic parameters $\theta_i$. In contrast, if presented with a finite number of CSs and their respective distances and prices, the human agent is very fast and efficient at making a choice. This observation guides us to apply a dual decomposition method to tackle problem (7). The independent variables of the relaxed problem (7) are $\xi_{ij}$ and $u_{ij}^t$, where $u_{ij}^t$ is relaxed to take continuous values:

$$u_{ij}^t \in [0, 1]. \quad (8)$$

A similar relaxation of $\xi_{ij}$ is not necessary as we will see shortly. Eq. (6) corresponds to $|N^d| \cdot |M^d|$ coupling constraints. We write the Lagrangian as:

$$\mathcal{L}(\xi_{ij}, u_{ij}^t, \lambda_i) = \sum_{i \in N^d} \sum_{j \in M^d} \left[ \xi_{ij} \cdot \theta_i \cdot (l_{ij})^2 + \lambda_{ij} \cdot \left( \left\lceil \frac{E_i}{e_j} \right\rceil \cdot \xi_{ij} - \sum_{t \in T^d} u_{ij}^t \right) \right].$$

The dual problem is indeed separable and can be decomposed into $|N^d| + |M^d|$ sub-problems. The first $|N^d|$ sub-problems are the ones where each user $i$ chooses the CS so as to minimize his/her disutility $Q_i$ coming from the CS's distance and prices (cost):

$$Q_i = \min_{\xi_{ij}} \left\{ \sum_{j \in M^d} \left[ \xi_{ij} \theta_i (l_{ij})^2 \right] + \sum_{j \in M^d} \left( \lambda_{ij} \cdot \left\lceil \frac{E_i}{e_j} \right\rceil \cdot \xi_{ij} \right) \right\}$$
$$\text{subject to } (1), (2). \quad (9)$$

Problem (9) is an integer program. However, because of constraint (2), it only has $|M^d|$ possible instances. This makes it very fast to solve, even by brute force methods. In practice, user $i$ would internally solve problem (9) by inspecting the locations and prices of the CS and making a choice. This fact keeps our formulation very much aligned with the realistic conceptualization of our setting.

The second group of sub-problems is the one where the CS schedules which EV is going to charge and which is going to wait, for all timeslots of the day, so as to make sure that each EV's SOC is at the desired level upon departure:

$$R_j = \max_{u_{ij}^t} \left\{ \sum_{i \in N^d} \sum_{t \in T^d} [\lambda_{ij} \cdot u_{ij}^t] \right\}$$
$$\text{subject to } (4), (5), (8) \quad (10)$$

The dual problem's objective is to compute prices $\lambda_{ij}$, so that the lower bound on the objective of problem (7) is maximized:

$$\max_{\lambda_{ij}} \{Q_i - R_j\}. \quad (11)$$

The coupling constraints (6) are satisfied by continuously updating the multipliers $\lambda_{ij}$ at each iteration $k$ of the algorithm. The intuitive interpretation is that $\lambda_{ij}$ is the price (bill) faced by EV $i$ at CS $j$. The price calculation is the job of an intermediary, namely the software behind the user's interface.

$$\lambda_{ij}^k = \lambda_{ij}^{k-1} + \varepsilon \cdot \max\left\{0, \left(\left\lceil \frac{E_i}{e_j} \right\rceil \cdot \xi_{ij} - \sum_{t \in T^d} u_{ij}^t \right)\right\}. \quad (12)$$

Usually the dual problem is tackled by running an iterative algorithm to converge to the optimal prices and then applying a rounding heuristic or treating the resulting continuous variables as probabilities for a choice of 1. However we need to keep contact with the realistic interpretation of our problem, where we need to provide the user with a definitive set of prices (one price for each CS, which will not change after the user selects) and let him make a one-off choice of his/her preferred CS, which the system will accept. Thus, the iterative method is not realistic for our setting. Nevertheless, the above formulation serves us in order to run offline simulations and stochastically determine the multipliers for the next day. The proposed method is outlined in the following section.

### III. STOCHASTIC APPROXIMATION OF THE OPTIMAL MULTIPLIERS

We model the number of users $|N^d|$ seeking to park their EV in the defined 2D area in a given day, as a Poisson random variable. We assume that each element of the user *type* $\psi_i$ as well as $C_j^t$ is drawn from an appropriate random distribution. The proposed system generates a number $|S|$ of scenarios $s \in S$ for the following day. Each scenario contains the number of EVs to arrive and their types:

$$s \triangleq \{|N^s|, \{\psi_i^s\}, i \in N^s\}.$$

We run day-ahead Monte Carlo simulations for the following day to obtain a distribution of the Langrage multipliers. Each multiplier depends on the CS's congestion as well as the (simulated) user's type. We then proceed to present heuristic methods of pre-determining each CS's asked payment from each EV upon arrival, based on the information we possess by both the offline simulations and the online parameters of the actual real-time system operation. We compare the approaches with a benchmark algorithm that is based on a first-come-first-serve approach.

### IV. REFERENCES

[1] E. Bitar and Y. Xu, "Deadline Differentiated Pricing of Deferrable Electric Loads," IEEE Trans. Smart Grid, 8(1), pp. 13-25, Jan. 2017.
[2] J. Escudero-Garzás and G. Seco-Granados, "Charging station selection optimization for plug-in electric vehicles: An oligopolistic game-theoretic framework," IEEE PES Innovative Smart Grid Technologies (ISGT), Washington, DC, 2012, pp. 1-8.
[3] E. Gerding, S. Stein, V. Robu, D. Zhao, and N. Jennings, "Two-sided online markets for electric vehicle charging," in Proc. Int. Conf. Auton. Agents Multi-Agent Syst., 2013, pp. 989–996.
[4] A. Koufakis, E. S. Rigas, N. Bassiliades and S. D. Ramchurn, "Towards an optimal EV charging scheduling scheme with V2G and V2V energy transfer," IEEE International Conference on Smart Grid Communications (SmartGridComm), Sydney, 2016, pp. 302-307